\documentclass[twocolumn,english,aps,eqsecnum]{revtex4-1}
\usepackage[T1]{fontenc}
\usepackage[latin9]{inputenc}
\usepackage{color}
\usepackage{bm}
\usepackage{amsmath}
\usepackage{amssymb}
\usepackage{graphicx}

\makeatletter
\usepackage{babel}

\makeatother

\usepackage{babel}
\begin{document}
\title{Retrocausal model of reality for quantum fields}
\author{Peter D. Drummond$^{1,2}$ and Margaret D. Reid$^{1,2}$}
\affiliation{$^{1}$ Centre for Quantum and Optical Science, Swinburne University
of Technology, Melbourne 3122, Australia}
\affiliation{$^{2}$ Weizmann Institute of Science, Rehovot 7610001 Israel}
\begin{abstract}
We show that one may interpret physical reality as random fields in
space-time. These have a probability given by the expectation of a
coherent state projection operator, called the Q-function. The resulting
dynamical evolution includes retrocausal effects. This suggests that
a physical universe exists without requiring observers, but with a
well-defined probability for its field configuration. By including
the meter dynamics, we show that field trajectories have quantum measurement
properties without wave-function collapse, including sharp measured
eigenvalues. We treat continuous and discrete measurements, and show
that this model predicts Bell inequality violations for measurements
on correlated spins. A discussion is give of a number of well-known
quantum paradoxes, showing how these can be treated in a realistic
model of measurement. Our theory resolves a number of practical and
philosophical issues in quantum measurement, and we compare it with
earlier theories.
\end{abstract}
\maketitle

\section{Introduction\label{sec:Introduction}}

The apparent quantum divide between macroscopic and microscopic worlds
has never been more important. Experimentalists now probe systems
of ever larger size \citep{brune1996observing,teufel2009nanomechanical},
yet more deeply quantum properties \citep{touzard2018coherent}. It
is commonplace for theoretical cosmologists to adopt the view of Coleman
\citep{Coleman1977,Callan1977} and others \citep{Liddle2000}, that
the entire universe is quantum mechanical. Such progress leaves the
paradoxes of the Copenhagen interpretation of quantum mechanics and
the Schr\"odinger cat \citep{schrodinger1935gegenwartige} in a central
position in physics. Yet, to use the standard postulates of measurement
theory \citep{dirac1981principles}, one must make a fundamental division
between the microscopic and macroscopic. This is not accepted by those
who see the universe as undivided \citep{BohmPhysRev.85.166}, and
is inconsistent with quantum cosmology.

Here we show that an ontological model of physical reality is obtainable
using a mapping to a phase-space of stochastic fields. The probability
of a configuration is given by the Q-function of quantum field theory.
As in statistical mechanics, any configuration has some probability
to occur, but only one exists. The theory includes vacuum fluctuations
and is frame invariant. This resolves paradoxes of quantum measurement
\citep{Einstein1935,schrodinger1935gegenwartige,bell1966problem},
since no separate observer and system is needed. We demonstrate that
the introduction of a model for the meter generates measurement predictions
identical to the usual ones.

The underlying fields provide an ontological model of both macroscopic
and microscopic realism \citep{leggett1985quantum,emary2013leggett,reid2017interpreting}.
This can be applied to the physical universe as a whole. No explicit
projection is needed to obtain sharply defined eigenvalues, although
one must include the physics of measurement. This eliminates the observer-dependent
wave-function collapse criticized by many physicists \citep{wigner1963problem,weinberg2006einstein}.
To demonstrate this, we study measurement models of discrete and continuous
measurements. We prove that Bell inequalities \citep{Bell1964} are
violated in this model of reality. 

In our approach, we accept the principle that measurement is central
to interpreting quantum theory. To measure a spin projection, one
must understand the polarizer. Until one does this, the spin wave-function
is an incomplete description. One can describe a system with or without
a meter. However, the results may change, as Bohr emphasized \citep{Bohr1935CanQuant}:
the meter is part of the universe.

Foundational problems in quantum mechanics have been the subject of
much study, with recent claims \citep{saunders2010many,harrigan2010einstein}
that the wave-function is ontological \citep{pusey2012reality,barrett2014no,ringbauer2015measurements}.
In these works, reality is defined as what is prepared in a laboratory,
with causality proceeding from past to future. 

We regard this as too narrow a definition. Physical reality surely
exists in all space-time, so that the most viable ontologies are space-time
fields. In our approach, stochastic fields in space-time are real
objects. This permits the theory to have a statistical interpretation
\citep{Leifer2014Quanta22,ringbauer2015measurements}. Theorems requiring
an ontological wave-function do not apply to our model \citep{ringbauer2015measurements}
because of retrocausality \citep{tetrode1922causal,dirac1938pam,wheeler1945interaction,pegg1982time,pegg1986absorber,pegg1999retrodiction}
due to negative diffusion terms in the dynamical equations \citep{altland2012quantum}.

Previous models include the de Broglie-Bohm and related theories with
both wave-function and particle coordinates \citep{BohmPhysRev.85.166,beltrametti1995classical},
discrete models used in quantum information \citep{aaronson2005quantum,spekkens2007evidence},
and a non-relativistic phase-space model with an epistemic restriction
\citep{budiyono2017quantum}. Other theories include consistent \citep{griffiths1984consistent}
or decoherent histories \citep{dieks1988formalism,dowker1992quantum,gell1993classical},
and quantum Darwinism \citep{Zurek_PhysRevA.73.062310}. These are
related to relative state or ``many-worlds'' theories \citep{Everett:1957}
which depend on observers.

By comparison, our model applies to quantum fields, is compatible
with relativity, and includes particle statistics. We define elements
of reality as fields in space-time, which exist at all times. These
satisfy an action principal with past and future boundary conditions.
This approach fulfills Einstein's requirements \citep{einstein1949albert}
that it is complete, formulated in terms of space/time fields, and
objective, without needing observers. In simple terms\textcolor{red}{,
}we generate a definite probability for obtaining one universe, which
gives a well-defined, objective procedure for measurement and quantum
models of the universe.

In our model, the only requirement to understand measurement is the
inclusion of a meter. Our use of a meter for this purpose follows
Bohr and Bell \citep{bell2004speakable,Bohr1935CanQuant}. No additional
decoherence mechanism \citep{ghirardi1986unified,diosi1987universal,penrose1996gravity}
or nonlinearity \citep{bialynicki1976nonlinear} is necessary to achieve
this. 

\section{A stochastic field reality model\label{sec:A-stochastic-field}}

The question of what is the reality that quantum theory describes
was raised by Bohr, Einstein, Born, and others. It was the subject
of much debate at the Solvay conferences \citep{bohr1996discussion}.
In this section a reality model for quantum theory is proposed using
stochastic fields.

\subsection{Definitions of reality}

The definition of reality most common in quantum mechanics uses Bohr's
interpretation \citep{Bohr1935CanQuant}. This claims that one can
only speak of reality in terms of macroscopic physical measurements.
To obtain measured results, the elements of microscopic reality, if
they exist, must be amplified to macroscopic levels. To quote Bohr
\citep{bohr1987essays}: ``the unambiguous account of proper quantum
phenomena must, in principle, include a description of all relevant
features of the experimental arrangement.''

In his Autobiographical Notes, Einstein \citep{einstein1949albert}
makes the following remark: ``There is no such thing as simultaneity
of distant events; consequently there is also no such thing as immediate
action at a distance in the sense of Newtonian mechanics.'' From
this, he concluded: ``It therefore appears unavoidable that physical
reality must be described in terms of continuous functions in space.'' 

This question of the existence of an objective reality was also raised
by Max Born, in his Nobel lecture \citep{born1955statistical}. In
view of difficulties in defining a simultaneous particle position
and momentum, Born asked ``what is the reality which our theory has
been invented to describe?''. He suggested that physical reality
might have a more abstract basis: ``For this, well-developed concepts
are available which appear in mathematics under the name of invariants
in transformations. Every object that we perceive appears in innumerable
aspects.''

If one accepts Einstein's viewpoint, then one should define reality
as an objective, continuous set of events, defined locally in space
and time. By objective, we mean that the events do not \emph{require}
an observer for their existence. However, according to Bohr, if observers
are present, one must not expect reality to be independent of them,
either. In observing a system, an interaction occurs which modifies
both system and observer.

We propose a definition of reality that unifies all three viewpoints.
Reality should exist in space and time, have a mathematical expression
in terms of invariants, and agree with quantum predictions of macroscopic
measurements. This does \emph{not} mean that reality ceases to exist
or is undefined in the absence of measurement. As elements of reality,
our approach uses stochastic fields which exist at all times, in the
future as in the past, in order to fulfill Einstein's requirements.
To calculate probabilities, we use the Q-function of quantum mechanics,
which is a positive and unique distribution that exists for all quantum
states. It is interpreted as the probability for a particular field
configuration defined at a given time. 

\subsection{The stochastic field probability}

We define the probability of a particular realization $\bm{\lambda}$
at time $t$ as the positive definite function $Q$, where:
\begin{equation}
Q\left(\bm{\lambda},t\right)=Tr\left[\hat{\rho}\left(t\right)\hat{\Lambda}\left(\bm{\lambda}\right)\right].\label{eq:Definition}
\end{equation}
The projector $\hat{\Lambda}$ is normalized such that $\int\hat{\Lambda}\left(\bm{\lambda}\right)d\bm{\lambda}=\hat{1},$
which means that the distribution is normalized so that 
\begin{equation}
\int Q\left(\bm{\lambda}\right)d\bm{\lambda}=1.
\end{equation}
 The configurations $\bm{\lambda}$ are representations of the Lie
groups compatible with the commutation relations, reflecting Born's
remark about invariants. We define $\bm{\lambda}=\left[\bm{\psi},\bm{\xi}\right]$
where $\bm{\psi}=\left[\psi_{1},\psi_{2}\dots\right]$ are fields
representing bosons. Since the physical universe comprises fermions
and bosons, we include a fermionic representation as well \citep{ResUnityFGO:2013,FermiQ}.
Here $\bm{\xi}=\left[\xi_{1},\xi_{2},\dots\right]$ are real antisymmetric
matrices representing fermions, and $\hat{\Lambda}\left(\bm{\lambda}\right)=\prod_{b,f}\hat{\Lambda}_{b}\left(\bm{\psi}_{b}\right)\hat{\Lambda}_{f}\left(\xi_{f}\right)$
is a Gaussian, positive-definite \citep{Glauber1963-states} operator
for bosons $(b)$ and fermions $(f)$. 

For simplicity, in this paper we construct Q-functions from coherent
state projectors. We first expand the operator field $\hat{\bm{\psi}}\left(\bm{r}\right)$
in annihilation and creation operators, $\hat{a}_{n}$ and $\hat{a}_{n}^{\dagger}$,
where $\left[\hat{a}_{n},\hat{a}_{m}^{\dagger}\right]=\delta_{nm}$.
The coherent state approach involves stochastic fields directly, since
in this case one can define:
\begin{equation}
\hat{\Lambda}_{b}\left(\bm{\psi}\right)=\frac{1}{\pi^{M}}\left|\bm{\alpha}\right\rangle _{c}\left\langle \bm{\alpha}\right|_{c}.
\end{equation}
Here $M$ is the number of spatial and spin modes in the bosonic field
expansion, and the coherent states $\left|\bm{\alpha}\right\rangle _{c}$
are defined \citep{Glauber1963-states} as follows:
\begin{equation}
\left|\bm{\alpha}\right\rangle _{c}=e^{\bm{\alpha}\cdot\hat{\bm{a}}^{\dagger}-\left|\bm{\alpha}\right|^{2}/2}\left|\bm{0}\right\rangle .
\end{equation}
The field $\bm{\psi}\left(\bm{r}\right)$ is obtained from $\hat{\bm{\psi}}\left(\bm{r}\right)$
by replacing the annihilation and creation operators operators with
complex amplitudes $\alpha_{n}$ and $\alpha_{n}^{*}$ respectively.
For a hermitian quantum field $\bm{\hat{\psi}}\left(\bm{r}\right)$,
the corresponding element of reality is
\begin{equation}
\bm{\psi}\left(\bm{r}\right)=\sum_{n=1}^{M}\left[\bm{u}_{n}\left(\bm{r}\right)\alpha_{n}+\bm{u}_{n}^{*}\left(\bm{r}\right)\alpha_{n}^{*}\right],
\end{equation}
where $\bm{u}_{n}\left(\bm{r}\right)$ are the normalized spatial
mode functions. This expansion is directly applicable to electromagnetic
modes, which will be used to illustrate the properties of measurements.
As a result, we will consider $\bm{\lambda}$ as comprising a set
of $M$ mode amplitudes $\bm{\alpha}$ in the examples described here.
These can be converted into $2M$ real quadrature amplitudes, $\bm{q}=\bm{\alpha}+\bm{\alpha}^{*}$
and $\bm{p}=\left(\bm{\alpha}-\bm{\alpha}^{*}\right)/i$.

Massive particles including fermions and bosons in the QCD Higgs sector
require more sophisticated methods involving Gaussian functions of
field operators, which are described elsewhere \citep{perelomov1972coherent,Corney_PD_PRL2004_GQMC_ferm_bos,Corney_PD_JPA_2006_GR_fermions,ResUnityFGO:2013,FermiQ,joseph2018phase}.
Antisymmetric matrices $\bm{\xi}$ can be transformed into a matrix
form equivalent to real and anomalous correlations of fermions. These
in turn can be reduced to products of space-time fields. A similar
method can be used for bosons with symmetric real matrices. This establishes
a link with the symmetry properties and invariants of the group-theoretic
homogeneous spaces of Cartan \citep{cartan1935domaines} and Hua \citep{Hua_Book_harmonic_analysis}. 

The Q-function is the probability at some time $t$, but in this model,
fields have continuous trajectories $\bm{\lambda}(t)$ defined at
all times. Before giving a detailed analysis, we note that our proposal
is not ruled out by no-go theorems \citep{montina2006condition,paavola2011finite}
for phase-space models, which have claimed that these cannot be realistic
theories. Such theorems do not allow for the presence of meters in
a measurement.

The necessity of including the meter in an analysis of macroscopic
measurements implies that a theory that leaves out the physics of
the meter is unrealistic. Omitting meters leads to relatively large
vacuum fluctuations. We show below that the inclusion of the meter
- regarded as an amplifier - eliminates this problem. Quantum fluctuations
become asymptotically negligible compared to the measured eigenvalue,
in the large gain limit. In this sense our model of reality is contextual
\citep{kochen1975problem}, since the meter changes what is measured
due to retrocausality.

\subsection{Uniqueness}

We next consider the question of uniqueness. There is no one-to-one
correspondence between a phase space variable and a quantum state.
A stochastic phase-space coordinate $\bm{\alpha}$ does not uniquely
define a quantum state, and nor does the quantum state uniquely define
$\bm{\alpha}$. Rather, a quantum state is defined by a distribution
$Q(\bm{\alpha})$ for $\bm{\alpha}$. If one regards quantum states
as elements of reality \citep{pusey2012reality}, then the lack of
a unique mapping between quantum states and the phase-space coordinates
used here may seem to be a problem. 

However, in any measurement of a quantum state, where the state is
not an eigenstate of the measurement operator, random results occur.
Since even random measurement results are real, one must have an explanation
of these. They could be thought to originate from the measurement
apparatus, but this gives an infinite recursion, where larger and
larger quantum states are required. For this reason, it is not unreasonable
to suppose that the quantum state is an incomplete description, just
as in the ensembles of statistical mechanics.

Our conclusion is that the originators of quantum mechanics regarded
the quantum wave-function as statistical for good reason. A one-to-one
mapping of states to reality is not necessary. There are many possible
phase-space variables that can occur in a given quantum state, consistent
with a statistical picture. This is consistent with the fact that
quantum measurements give random results at the level of Planck's
constant, apart from the case where an eigenstate is prepared. Even
this may have microscopic randomness, as we show later.

\begin{figure}
\begin{centering}
\includegraphics[width=1\columnwidth]{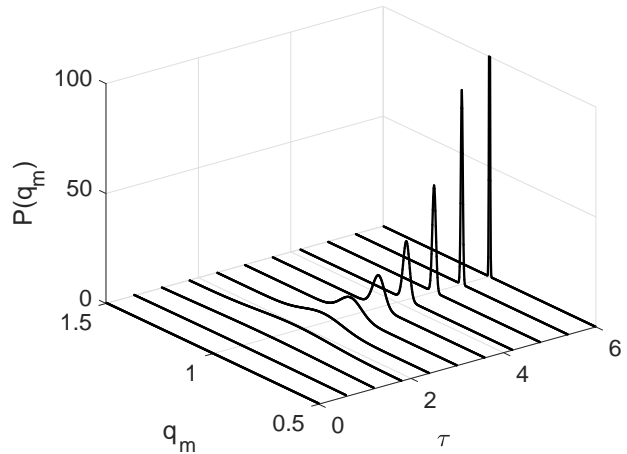}
\par\end{centering}
\caption{Probability density $P\left(q_{m}\right)$ as a function of the measured
quadrature amplitude $q_{m}$ and dimensionless time $\tau=gt$ for
a $q$ eigenstate with $q_{0}=1$. The increasing sharpness with gain
shows how a continuous eigenvalue can be recovered despite vacuum
noise. The vertical scale is truncated at $P\left(q_{m}\right)=100$
for visibility. \label{fig:Inferred-q-distribution}}
\end{figure}

There is a second question concerning uniqueness. In the single-mode
case one may consider an eigenstate for an observable $\hat{q}=\hat{a}+\hat{a}^{\dagger}$,
with eigenvalue $q_{1}$. Another eigenstate could have eigenvalue
$q_{2}$. Because of the width of the Q function for each eigenstate,
the same $\alpha$ could correspond to \emph{either} outcome for $\hat{q}.$
A given $\alpha$ therefore does \emph{not} correspond to a unique
eigenvalue for a measurement outcome. We explain below how this is
consistent with the requirement of time-symmetric boundary conditions,
which give unique measurement outcomes for eigenstates, as shown in
Fig (\ref{fig:Inferred-q-distribution}).

\subsection{Alternative phase-space representatives}

There are many quasi-probabilities for quantum fields, in both bosonic
and fermionic cases: why should one choose the Q-function? The Wigner
representation \citep{Wigner1932} was proposed as a statistical theory
by Moyal \citep{moyal1949quantum}, but is non-positive, and has no
direct probabilistic interpretation \citep{dirac1945analogy,Feynman1982}.
Another possibility, the Glauber P-function \citep{Glauber1963-states},
is non-positive and highly singular for some quantum states. There
is a continuous class of `s-ordered' bosonic quasi-probabilities \citep{Cahill1969}.
The case of $s=-1$ is the Q-function, which is the most compact of
the s-ordered distributions that are always positive. The positive
P-function has a distribution which is positive and non-singular.
Since it is not unique \citep{Drummond1980}, it does not give a distinctive
model.

In the fermionic case, there is a positive Q-function which we use.
There are also Grassman-valued P-distribution quasi-probabilities
\citep{cahill1999density}. However, these have no probabilistic interpretation.
As in the bosonic case, there are P-distributions using expansions
in fermionic Gaussian states. These are either not unique, \citep{Corney_PD_JPA_2006_GR_fermions,Corney_PD_PRB_2006_GPSR_fermions,Corney_PD_PRL2004_GQMC_ferm_bos}
or else not always positive \citep{Riashock2018}. 

In summary, in our approach to the measurement problem we require
that the theory has predictions equivalent to quantum mechanics. It
should be unique, and have a positive probability distribution. Each
possible real outcome that can occur must have a well-defined dynamical
equation. A single trajectory of stochastic fields defined in space-time
then corresponds to a possible universe. This resembles the way that
statistical mechanics describes an ensemble: There are many possible
outcomes, but only one is realized. 

The generalized Q-function distribution for stochastic fields \citep{Husimi1940,FermiQ}
satisfies our requirements for an ontology underlying quantum mechanics.
Other phase-space representations, like the Wigner representation,
are either non-positive or non-unique.  We will show that the fact
that the Q-function phase-space trajectory is determined by future
as well as past boundary conditions addresses issues associated with
non-orthogonality of coherent states. This results in measurement
outcomes that correspond to experiment.

\section{Stochastic field dynamics\label{sec:Stochastic-field-dynamics}}

Q function distributions also include vacuum fluctuations. We regard
these as real, since the fields are ontological. From the Heisenberg
uncertainty principle, the Q-function is not infinitely sharp, leading
to restrictions on physically valid distributions. Yet eigenvalues
of hermitian operators do exist. In fact, these sharply defined outcomes
are fundamental to quantum mechanics. To understand this issue, we
analyze measurements.

To treat measurement dynamics, one can make use of several techniques
for analyzing Q-function time-evolution. Distributions can be calculated
using the time-evolution of $\hat{\rho}\left(t\right)$, which follows
standard techniques. It is also instructive to follow individual trajectories
as realistic dynamical histories. Here, we present a summary of this
issue, by remarking that the probability distribution obeys a first
order differential equation in time. This dynamical equation is obtained
uniquely and exactly from equivalence theorems and identities that
map quantum operators into differential operators.

\subsection{Generalized Fokker-Planck equations}

Unitary dynamics is given in quantum field theory by the standard
result that $i\hbar\partial_{t}\hat{\rho}=\left[\hat{H},\hat{\rho}\right]$,
where we use the notation that $\partial_{t}\equiv\partial/\partial t$.
This implies that the probability distribution $Q\left(\bm{\lambda},t\right)$
evolves according to
\begin{equation}
\partial_{t}Q\left(\bm{\lambda},t\right)=\frac{i}{\hbar}Tr\left\{ \left[\hat{H},\hat{\Lambda}\left(\bm{\lambda}\right)\right]\hat{\rho}\left(t\right)\right\} .
\end{equation}
Operator expressions occurring inside the trace, of form $\hat{H}\hat{\Lambda}$
or $\hat{\Lambda}\hat{H}$, can then be reduced to differential expressions
in the phase space. In the single-mode bosonic case, defining $\partial_{\alpha}\equiv\partial/\partial\alpha$,
the following differential identities are well known:
\begin{eqnarray}
\hat{a}^{\dagger}\left|\alpha\right\rangle _{c}\left\langle \alpha\right|_{c} & = & \left[\partial_{\alpha}+\alpha^{*}\right]\left|\alpha\right\rangle _{c}\left\langle \alpha\right|_{c}\nonumber \\
\hat{a}\left|\alpha\right\rangle _{c}\left\langle \alpha\right|_{c} & = & \alpha\left|\alpha\right\rangle _{c}\left\langle \alpha\right|_{c}.\label{eq:identities-1-2}
\end{eqnarray}

Examples of these mappings to obtain unitary Q-function dynamics are
given in the next section. Using these and other operator identities
as required, this is equivalent to a zero-trace generalized diffusion
equation for the elements of reality $\bm{\lambda}$, of form
\begin{equation}
\dot{Q}\left(\bm{\lambda},t\right)=\mathcal{L}\left(\bm{\lambda}\right)Q\left(\bm{\lambda},t\right).
\end{equation}
The resulting differential equation has a diffusive character if the
differential terms are no higher than second order. Such terms are
found in the case of Hamiltonian evolution with no more than quartic
nonlinearities in the fields. This covers the most important fundamental
quantum field theories, including quantum electrodynamics. This algebraic
restriction is compatible with quantum field Hamiltonians, but not
particle Hamiltonians in first quantization. Therefore, these methods
apply to quantum fields, not quantum particles. 

The dynamical equation for $Q$ is a generalized Fokker-Planck equation.
It has an action principle with both past and future boundary conditions
\citep{drummond2019time}. The dynamics is equivalent to quantum mechanics,
so it is consistent with known physical observations. Generalized
Fokker-Planck equations are not the same as the dissipative equations
that describe coupling to reservoirs. In particular, they have a zero-trace
diffusion matrix for unitary evolution, which leads to a retrocausal
interpretation.

Although there are similarities with dissipative equations, unitary
evolution under a Hamiltonian is \emph{reversible}. The corresponding
Q-function dynamical equations have well-defined solutions whether
boundary conditions are specified in the past or the future. This
occurs because of second order derivative terms which are not positive
definite. The dynamical equations for the distribution are therefore
not Fokker-Planck equations. There is also no conventional stochastic
equation to describe an individual trajectory.

This issue of the existence of a stochastic equation is immaterial
if one regards the entire distribution as a single element of reality.
From this perspective, an individual phase-space coordinate, whether
a field or a complex amplitude, is only a part of the wave-function
or distribution. However, since we take the view that an \emph{individual}
stochastic field is an element of reality, its trajectory is important.
This satisfies a real action principle equivalent to a forwards-backwards
stochastic equation, propagating in \emph{both }time directions.

To give more detail, consider a generalized Fokker-Planck dynamics.
Without loss of generality, we can use real phase-space variables
$\phi^{\mu}$ such that
\begin{equation}
\dot{Q}=\partial_{\mu}\left[-A^{\mu}+\frac{1}{2}\partial_{\nu}D^{\mu\nu}\right]Q.\label{eq:generalized FPE}
\end{equation}
For simplicity, we assume that $D^{\mu\nu}=D^{\mu}\delta^{\mu\nu}$
is diagonal and constant, which is valid for the amplifier examples
used here. For $M$ modes in bosonic cases, the $2M$ variables are
divided into positive and negative diffusion terms, $\bm{\phi}=\left[\bm{p},\bm{q}\right],$
such that if $\mu\le M$ then $D^{\mu}>0$; otherwise $D^{\mu}<0$.
The $q$ and $p$ are real, complementary field quadrature amplitudes,
not mechanical position and momenta coordinates.

By contrast, a conventional dissipative Fokker-Planck equation has
$D^{\mu}>0$ for all $\mu$ \citep{graham1977path}. This is typically
obtained by including a large number of reservoir degrees of freedom,
together with other approximations. We emphasize that there is no
reservoir required here. The present result is obtained directly from
the Hamiltonian, and there is no dissipation. Hence, this method is
fundamental, provided there are no higher order nonlinearities than
quartic. Dissipative reservoirs can also be included if required.

There is a corresponding real path integral that defines the possible
trajectories of the solutions, which is:
\begin{equation}
Q\left(\bm{\phi}_{f},t_{f}\left|\bm{\phi}_{0},t_{0}\right.\right)\propto e^{-\int\mathcal{L}\left(\bm{\phi},\dot{\bm{\phi}}\right)dt}.
\end{equation}
This is different to a Feynman path integral which gives complex amplitudes,
with paths that have no realistic interpretation. Here the Lagrangian
is real, and there is no imaginary coefficient. For constant diffusion,
the Lagrangian density has the form
\begin{equation}
\mathcal{L}=\frac{1}{2}\sum_{\mu}\left\{ \frac{1}{d^{\mu}}\left(\dot{\phi}^{\mu}-A^{\mu}\right)^{2}-s_{\mu}\partial_{\mu}A^{\mu}\right\} ,
\end{equation}
where $d^{\mu}=\left|D^{\mu}\right|$ and $s_{\mu}=Sign(D^{\mu})$.
The path integral boundary conditions for $\bm{p}$ are fixed in the
past and open in the future, while those for $\bm{x}$ are fixed in
the future and open in the past. 

This is equivalent to a forwards-backwards stochastic equation , with
gaussian delta-correlated noises $\left[\bm{w}_{p},\bm{w}_{q}\right]$
such that $\left\langle \dot{w}^{\mu}\left(t\right)\dot{w}^{\nu}\left(t'\right)\right\rangle =\left|D^{\mu\nu}\right|\delta\left(t-t'\right)$.
The relevant equations are as follows: 
\begin{align}
\bm{p}(t) & =\bm{p}(t_{0})+\int_{t_{0}}^{t}\bm{A}_{p}(t')dt'+\int_{t_{0}}^{t}d\bm{w}_{p}\nonumber \\
\bm{q}(t) & =\bm{q}(t_{f})-\int_{t}^{t_{f}}\bm{A}_{q}(t')dt'-\int_{t}^{t_{f}}d\bm{w}_{q}\,.\label{eq:forward-backward}
\end{align}

Q-functions are real probability distributions, and one can interpret
an individual path as a possible stochastic trajectory. From Eq (\ref{eq:forward-backward}),
trajectory boundary values must be specified \emph{both} in the past
\emph{and} in the future, depending on the quadrature, which is different
to having initial conditions only. 

\subsection{Boundary conditions }

The boundary conditions are uniquely constrained by the quantum state.
These distributions, specified in the past and the future, must lead
dynamically to the initial distribution $Q\left(\bm{\phi},t_{0}\right)$,
which is related through Eq (\ref{eq:Definition}) to the required
initial quantum state. This is a subtle issue, and we describe it
in detail in this subsection.

In quantum mechanics, the assumption is generally made that any quantum
state can be generated on demand. For simplicity, we make this assumption
also, and in this respect our approach is like the usual treatment.
In practice, experimental state preparation is nontrivial. Conditional
state preparation is often necessary when high fidelities are required
\citep{liang2019quantum}. Thus, there are quantum states where preparation
on demand is far from easy, and this requires discussion, given the
approach to reality described here.

For example, a common way to prepare a specific field quadrature is
through an EPR experiment \citep{Reid:1989} in which two correlated
quadratures are prepared through down-conversion. The specification
of the eigenvalue may not be known until a future time measurement
is made of the correlated quadrature. This can be made in the future,
and viewed as a future-time boundary value. A careful analysis suggests
that this is not uncommon. Future boundary conditions are implicit
in the analysis of Maxwell's equations for radiating fields \citep{dirac1938pam}.
They are discussed for both classical \citep{wheeler1945interaction}
and quantum fields \citep{pegg1986absorber}, and are as fundamental
as past time boundary conditions.

The existence of a quantum state or density matrix $\hat{\rho}\left(t_{0}\right)$
at a given time $t_{0}$ is equivalent to having a well-defined distribution
$Q\left(\bm{\phi},t_{0}\right)$. Yet elements of reality must exist
both in the past and future, so that they are defined in all reference
frames. Given the equations above, the dynamics of any phase-space
coordinate $\bm{\phi}$ requires probabilities of initial conditions
for forward-time components $\bm{p}$, and probabilities of final
conditions for components $\bm{x}$ that propagate to the past. 

This approach appears to differ from everyday experience. A common
philosophy is that causality must be uni-directional, but this is
essentially a classical world-view. It is usually expected that one
may prepare any required quantum state, which propagates to the future.
However, retrocausal dynamics requires the specification of both initial
and final conditions. To obtain a given $\hat{\rho}\left(t_{0}\right)$,
these boundary conditions on $\bm{\phi}$ and their distributions
in the future and past must therefore dynamically give rise to $Q\left(\bm{\phi},t_{0}\right)$
at the required time $t_{0}$. 

In summary, an element of reality requires the specification of both
future and past time boundary values for its dynamics. 

\section{Stochastic measurement models \label{sec:Idealized-models-of}}

In quantum mechanics, it is a standard postulate that any state $\hat{\rho}$
can be experimentally prepared, and any hermitian operator $\hat{O}$
can be measured. However, Bohr was careful to insist that in the case
of measurement, one has to consider the experimental apparatus used
to obtain the macroscopic output measured \citep{Bohr1935CanQuant,bohr1996discussion}.
State preparation and measurement are complementary, and state preparation
therefore requires an operational definition as well, as discussed
above.

Here we will focus on the operational measurement issue, by considering
an example of a single bosonic mode, with $\alpha=x+ip$. The normalized
projector, $\hat{\Lambda}_{b}$ is:
\begin{equation}
\hat{\Lambda}_{b}=\frac{1}{\pi}\left|\alpha\right\rangle _{c}\left\langle \alpha\right|_{c}.\label{eq:bosonicLambda}
\end{equation}
For example, in a vacuum state $\left|0\right\rangle $, the distribution
has a finite variance due to vacuum fluctuations, given by
\begin{equation}
Q\left(\alpha\right)=\frac{1}{\pi}e^{-\left|\alpha\right|^{2}}.
\end{equation}

Yet it is not just the vacuum state that has a finite variance. In
fact, all states have substantial quantum fluctuations in this approach.
The most puzzling question to be answered with a stochastic field
interpretation of reality is therefore \emph{what this implies in
measurement terms.} How can real quantum fluctuations be compatible
with the fact that there are quantum states with well-defined eigenvalues? 

\subsection{Measurement of a continuous eigenvalue}

To understand this, we will treat the theory of measurement in an
idealized case. For simplicity, we expand the quantum fields in mode
operators. We first treat a model for the measurement of a continuous
variable, the $\hat{q}$ quadrature of the single-mode radiation field.
Here, we consider the simplest theory for a measurement of $\hat{x}$,
corresponding to a measurement on the system by a meter, where the
system observable of interest is directly amplified using a low-noise
parametric amplifier \citep{Yuen1976,walls1983squeezed,drummond2014quantum}.
These are widely used practical devices in modern quantum technology
\citep{Slusher:87,VahlbruckSchnabel2016PhysRevLett.117.110801,MalnouLehnert2018PhysRevApplied.9.044023},
and epitomize how one makes quantum-limited field quadrature measurements
in the laboratory. Such technologies are also used for quantum limited
position measurements \citep{caves1981quantum,braginsky_khalili_thorne_1992,power1997scheme}. 

We let $\hat{q}=\hat{a}+\hat{a}^{\dagger}$ and $\hat{p}=\left(\hat{a}-\hat{a}^{\dagger}\right)/i$
be the field quadratures of mode $\hat{a}$. The amplification takes
place via interaction of the system with a meter. In our model, the
interaction Hamiltonian in the rotating frame of our meter is that
given by a parametric amplifier: 
\begin{equation}
\widehat{H}=\frac{i\hbar g}{2}\left[\hat{a}^{\dagger2}-\hat{a}^{2}\right]\,.
\end{equation}
We can obtain the evolution of the system and meter by solving\textcolor{red}{{}
}for the quantum state or by using the Heisenberg equations, which
have operator solutions with $\hat{q}\left(t\right)=\hat{q}\left(0\right)e^{gt}$
and $\hat{p}\left(t\right)=\hat{p}\left(0\right)e^{-gt}$. For the
choice $g>0$, we see that $\hat{q}$ is amplified, as required if
we are to view the meter as a measuring device for $\hat{x}$. Equivalently,
it is also possible to use the Q-function Fokker-Planck or stochastic
equations, which we\textcolor{red}{{} }explain below. Given an initial
vacuum state, in which $\left\langle \Delta\hat{q}^{2}\left(0\right)\right\rangle =\left\langle \Delta\hat{p}^{2}\left(0\right)\right\rangle =1$,
the $\hat{p}$ quadrature becomes squeezed as time evolves, with the
variance $\left\langle \Delta\hat{p}^{2}\left(t\right)\right\rangle $
reducing below the vacuum level, while the $\hat{q}$ quadrature develops
a large variance.

Suppose the quantum system is prepared in a superposition of eigenstates
$\left|q_{0}\right\rangle _{x}$ of the $\hat{q}$ quadrature, each
with eigenvalue $q_{0},$ and a total variance $\left\langle \Delta\hat{q}^{2}\left(0\right)\right\rangle $.
The parametric amplifier amplifies the $\hat{q}$ quadrature to a
macroscopic level. After measurement, if the gain is $G\left(\tau\right)=e^{\tau}$,
where $\tau\equiv gt$ is the dimensionless time, the resulting variances
in the phase-space variables are: 
\begin{align}
\left\langle \Delta q^{2}(t)\right\rangle  & =1+G^{2}\left\langle \Delta\hat{q}^{2}\left(0\right)\right\rangle \nonumber \\
\left\langle \Delta p^{2}(t)\right\rangle  & =1+\left\langle \Delta\hat{p}^{2}\left(0\right)\right\rangle /G^{2}.\label{eq:measured variance}
\end{align}
The vacuum noise contribution to the ontological phase-space variable
$q\left(\tau\right)$ has a variance of unity, even in an eigenstate
of $\hat{q}$ where $\left\langle \Delta\hat{q}^{2}\left(0\right)\right\rangle =0$.
Yet experimentalists can identify a reproducible eigenvalue $q_{0}$
after the measurement. This is possible, because the measured estimate
of the eigenvalue is $q_{m}=q(\tau)/G(\tau).$ The variance in $q(\tau)$
- regarded as a real event - is now dominated by $\left\langle \Delta\hat{q}^{2}\left(0\right)\right\rangle $.
After measurement, the phase-space variable $q$ is the actual \emph{result}
of measuring the $\hat{q}$ quadrature, once the gain $G$ is accounted
for. The measured value $q_{m}$ has a variance of $1/G^{2}$ that
vanishes in the asymptotic, large gain limit.

In the limit of an initial eigenstate, $\left|\psi\left(0\right)\right\rangle =\left|q_{0}\right\rangle _{x}$,
we find using standard techniques from quantum theory that after measurement
with dimensionless interaction time $\tau$, the marginal probability
for a Q-function trajectory with measured eigenvalue $q_{m}=q\left(\tau\right)/G\left(\tau\right)$
is:
\begin{equation}
P\left(q_{m},\tau\right)=\sqrt{\frac{G^{2}\left(\tau\right)}{2\pi}}exp\left(-\frac{1}{2}G^{2}\left(\tau\right)\left(q_{m}-q_{0}\right)^{2}\right).
\end{equation}

This probability distribution is plotted in Fig (\ref{fig:Inferred-q-distribution}),
demonstrating that the amplifier has the desired effect. It causes
a relative narrowing in the probability distribution of measured outcomes.
There is no sharp boundary imposed between 'classical' and 'quantum'
worlds, no measurement projection, and not even any decoherence. Yet
the outcome is rather clear. The effect of a quantum limited amplifier
is to eliminate noise, leading to a well-defined outcome.

\subsection{Q-function dynamical equations}

The calculation given above used a standard quantum operator approach.
One can calculate the result of the measurement more directly by using
the phase-space dynamical trajectories. We now consider the same measurement
process using the Q-function dynamical equation,

\begin{equation}
\frac{dQ}{dt}=-\frac{g}{2}Tr\left\{ \left[\hat{a}^{\dagger2}-a^{2},\hat{\Lambda}\left(\bm{\alpha}\right)\right]\hat{\rho}\right\} .
\end{equation}
This can be written in terms of mode creation and annihilation operators
using the identities given in Eq (\ref{eq:identities-1-2}). After
making the substitutions of differential identities instead of operators,
the dynamical evolution of the Q-function for unitary evolution is
obtainable from the differential equation:
\begin{equation}
\frac{dQ}{dt}=-\frac{g}{2}\left[\frac{\partial^{2}}{\partial\alpha^{2}}+2\frac{\partial}{\partial\alpha}\alpha^{*}+hc\right]Q.
\end{equation}
We now make a variable change to real variables, so that $\alpha=x+ip$,
 which gives:
\begin{equation}
\frac{dQ}{d\tau}=\left[\frac{\partial}{\partial p}p+\frac{\partial^{2}}{\partial p^{2}}-\left(\frac{\partial}{\partial q}q+\frac{\partial^{2}}{\partial q^{2}}\right)\right]Q\,.
\end{equation}
The Q-function equation for $q$ combines gain with negative diffusion,
while that for $p$ is the exact time reversal or complement, since
it combines loss with positive diffusion. This leads to time-reversed
causation for the $q$ quadrature. The retrocausal evolution of the
measured variable gives insight into why the vacuum fluctuations associated
with the ontological phase-space variable $q$ for the $\hat{q}$
eigenstate, remain small during measurement of $q$. One might have
expected these vacuum fluctuations to grow due to amplification, but
this in fact does not happen, as shown by the trajectory graphs given
in Fig (\ref{fig:trajectories-amplified-eigenstate-x}).

A simple way to explain this is that an eigenstate has the minimum
possible level of quantum fluctuations in a stochastic field ontology.
This noise level is not zero, as one might expect from the Heisenberg
uncertainty principle. Instead, the Q-function noise remains at the
level of Planck's constant, as shown by Eq (\ref{eq:measured variance}).
The balance of noise and loss is such as to exactly maintain this
level of quantum noise in the limit of a highly squeezed state. This
behavior is similar to what occurs with gain, since gain is just time-reversed
quantum squeezing.

However, this residual vacuum noise is not observable in macroscopic
measurements. A measurement does not eliminate quantum noise, but
rather reduces it to an arbitrarily low level relative to the observable.
In a retrocausally amplified signal the \emph{relative} noise becomes
smaller with increased gain. This is simply the time-reversed corollary
of the well-known physics of causally attenuated signals, where relative
noise becomes larger with increased loss. The eigenvalue is obtained
as the asymptotic limit of a high-gain measurement. This explains
the distinction between microscopic and macroscopic behavior as a
limiting process. There is no artificial boundary to be crossed.

Such retro-causal behavior is known to also provide a mechanism for
overcoming the restrictions of Bell's theorem \citep{pegg1986absorber,wharton2019bell},
and the details of how this works in the present approach are given
in the next section.

To summarize these results, the Q-function stochastic field quadrature
after a measurement is $q=Gq_{0}+\epsilon$, where $\epsilon$ is
some random vacuum noise with $\left\langle \epsilon^{2}\right\rangle =1$.
From the amplified macroscopic value $x$, the experimentalist infers
a measured eigenvalue of $q_{m}=q_{0}+\epsilon/G$. A sharp eigenvalue
$q_{0}$ is recovered from the measured data in the limit of an ideal,
infinite gain meter, as $\tau\rightarrow\infty$, with no assumptions.
The Q-function dynamical equations give the same prediction as we
obtained with a standard quantum approach. One can also treat less
idealized measurements with noise and decoherence, if required.

\subsection{Trajectory picture}

For unitary Q-function evolution equations, the diffusion matrix is
traceless and equally divided into positive and negative definite
parts. In this case the $p$ quadrature decays and has positive diffusion,
while the the $x$ quadrature shows growth and amplification, but
has negative diffusion in the forward time direction. The amplified
quadrature, which corresponds to the measured signal of a parametric
amplifier, propagates stochastically backward in time from the future.

\begin{figure}
\begin{centering}
\includegraphics[width=1\columnwidth]{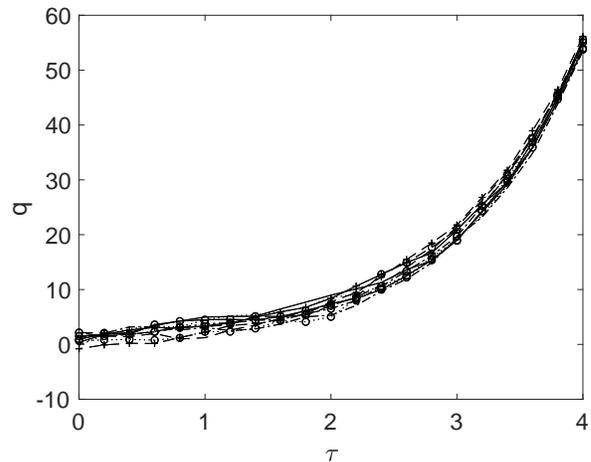}
\par\end{centering}
\caption{Graph of $10$ random trajectories of $q(\tau)$ vs. dimensionless
time $\tau$, for retrocausal evolution under the amplification due
to measurement of $\hat{x}$ on a system prepared in an eigenstate
$|q_{0}\rangle$ of $\hat{q}$ at $\tau=0$. The dynamics for $q$
are solved backward in time, starting with the final condition $q(\tau_{f})$.
The trajectories for $q$ correspond to a near eigenstate of $\hat{q}$.
The ensemble averaged noise at $\tau=0$ is the same as the final
noise at $\tau=\tau_{f}$, with a variance in $q$ of $1$. \label{fig:trajectories-amplified-eigenstate-x}}
\end{figure}

To obtain a single, probabilistic trajectory, one must therefore constrain
$x$ by a \emph{future} boundary condition. For simplicity, we suppose
the initial Q-function is factorizable. This is the case, for example,
if the eigenstate is modeled as a Gaussian squeezed state in the limit
of large quantum squeezing. Hence, the $Q$-function solutions can
always be factorized as a product with $Q\left(\alpha,\tau\right)=P_{q}\left(q,\tau\right)P_{p}\left(p,\tau\right)$.
Then, if $\tau_{-}=\tau_{f}-\tau$ is the backward time direction,
the time-evolution of each of these factors is quasi-diffusive, with
an identical diffusive structure, except that it occurs in each of
two \emph{different} time directions: 
\begin{align}
\frac{dP_{q}}{d\tau_{-}} & =\left[\frac{\partial}{\partial q}q+\frac{\partial^{2}}{\partial q^{2}}\right]P_{x}\nonumber \\
\frac{dP_{p}}{d\tau} & =\left[\frac{\partial}{\partial p}p+\frac{\partial^{2}}{\partial p^{2}}\right]P_{p}\,.
\end{align}
One can readily verify, by substitution, that the marginal distributions
of the Q-function are the unique solutions to these equations. These
are equivalent to the stochastic equations
\begin{align}
\frac{dq}{d\tau_{-}} & =-q+dw_{1}\nonumber \\
\frac{dp}{d\tau} & =-p+dw_{2}\,,
\end{align}
where the noise terms have correlations given by:
\begin{equation}
\left\langle dw_{\mu}\left(\tau\right)dw_{\nu}\left(\tau'\right)\right\rangle =2\delta_{\mu\nu}\delta\left(\tau-\tau'\right).
\end{equation}

\begin{figure}
\begin{centering}
\includegraphics[width=1\columnwidth]{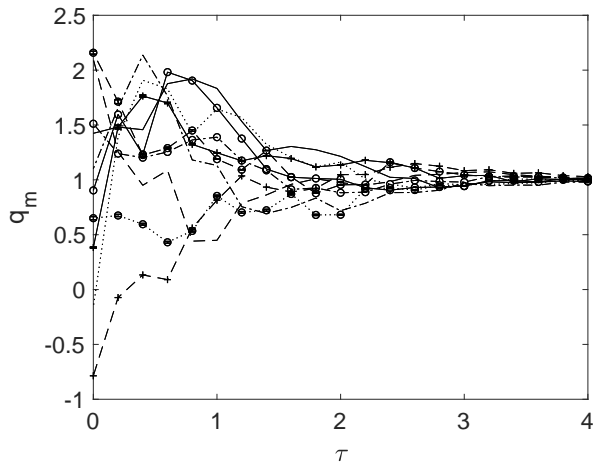}
\par\end{centering}
\caption{Graph of $10$ individual trajectories of the measured eigenvalue
$q_{m}\left(\tau\right)=q\left(\tau\right)/G\left(\tau\right)$ vs.
dimensionless time $\tau$, for evolution under the amplification
due to measurement of $\hat{q}$. The initial state is an eigenstate
$|q_{0}\rangle$ of $\hat{q}$. These are the same trajectories as
given in Fig (\ref{fig:trajectories-amplified-eigenstate-x}), showing
a decreasing relative variance with time $\tau$ and total gain $G$.
\label{fig:trajectories-measured eigenvalue}}
\end{figure}

Typical trajectory solutions to the stochastic equations for $q(\tau)$
with $x_{0}=1$ are given in Fig (\ref{fig:trajectories-amplified-eigenstate-x})
as a function of dimensionless time. The corresponding results for
$q_{m}=q/G$ are given in Fig (\ref{fig:trajectories-measured eigenvalue}),
showing how the inferred or measured result $q_{m}$ converge to the
exact eigenvalue, $q_{0}=1$, as the total gain increases with time.
These results agree with the distributions in Fig (\ref{fig:Inferred-q-distribution})
for the marginal distributions of $q_{m}$, calculated either using
the pure state solutions $\hat{\rho}(t)$ or the Q-function equations.

\subsection{Measurement of a discrete eigenvalue}

We next consider measurement of a discrete eigenvalue, as typified
by a two level system or qubit. Suppose the qubit is in a superposition
state, $\left|\psi\right\rangle _{q}=\left(\left|\uparrow\right\rangle +\left|\downarrow\right\rangle \right)/\sqrt{2}.$
This is equivalent to a spin $1/2$ state, with a Pauli spin operator
$\hat{\sigma}_{z}=\left|\uparrow\right\rangle \left\langle \uparrow\right|-\left|\downarrow\right\rangle \left\langle \downarrow\right|$.
The outcome of measuring $\hat{\sigma}_{z}$ is either $1$ or $-1$.
To model the measurement of $\hat{\sigma}_{z}$, we employ a commonly
used experimental Hamiltonian \citep{rosales2018weak,blais2004cavity,wallraff2004strong}:
\begin{equation}
H_{M}=\frac{\hbar g}{2}\hat{n}\hat{\sigma}_{z}\,.\label{eq:ham}
\end{equation}
The measurement is performed by coupling the qubit to an optical field.
The field is a single mode with boson operator $\hat{a}$ and number
operator $\hat{n}=\hat{a}^{\dagger}\hat{a}$. The optical meter field
is prepared in a coherent state $\left|G/i\right\rangle _{c}$ and
coupled for a time $t_{m}$. Thus, the coupled initial state is:
\begin{equation}
\left|\psi\left(0\right)\right\rangle _{qm}=\left(\left|\uparrow\right\rangle +\left|\downarrow\right\rangle \right)\left|G/i\right\rangle _{c}/\sqrt{2}.
\end{equation}
Here $G$ is the gain of the meter. This approach is even applicable
in the limit of $G\rightarrow0$, where it yields the large fluctuations
found in weak measurement theory \citep{Aharonov1988,rosales2018weak}.
However, we are interested in the macroscopic or large $G$ limit,
so that vacuum fluctuations are relatively negligible. With an input
superposition incident on the measurement device, the final measured
state after evolution for a dimensionless measurement time of $\tau_{m}=\pi$,
where $\tau=gt$, is: 
\begin{eqnarray}
\left|\psi\left(\tau_{m}\right)\right\rangle _{qm} & = & \frac{1}{\sqrt{2}}\left[\left|\uparrow\right\rangle \left|G\right\rangle _{c}+\left|\downarrow\right\rangle \left|-G\right\rangle _{c}\right]\,.\label{eq:corr-cat-2}
\end{eqnarray}
This describes the final state of the two-mode system and the final
state of the meter field. An observation is then made of the real
quadrature $x$ of the meter field, regarded as an element of reality. 

Here the stochastic description includes a spin degree of freedom,
which can be treated either using a fermionic \citep{FermiQ} or SU(2)
atomic coherent state \citep{Arecchi1972,altland2012quantum} Q-function.
The simplest method uses the SU(2) atomic coherent state $\left|z\right\rangle _{a}$,
defined as
\begin{equation}
\left|z\right\rangle _{a}=\frac{1}{\sqrt{1+\left|z\right|^{2}}}\exp\left(z\hat{\sigma}^{+}\right)\left|\downarrow\right\rangle ,
\end{equation}
where $\hat{\sigma}^{+}=\left|\uparrow\right\rangle \left\langle \downarrow\right|$
is the qubit raising operator. The normalized basis for the spin Q-function
is then:
\begin{equation}
\hat{\Lambda}_{f}\left(z\right)=\frac{2}{\pi\left(1+zz^{*}\right)^{2}}\left|z\right\rangle \left\langle z\right|_{a}.
\end{equation}
In this case, the coupled Q-function is given by:
\begin{equation}
Q\left(\alpha,z\right)=Tr\left[\hat{\rho}\hat{\Lambda}_{f}\left(z\right)\hat{\Lambda}_{b}\left(\alpha\right)\right].
\end{equation}

The spin differential identity for the Pauli spin operator z-component
is:
\begin{align}
\hat{\sigma}_{z}\left|z\right\rangle \left\langle z\right|_{a} & =\left[2z\partial_{z}+\frac{zz^{*}-1}{1+zz^{*}}\right]\left|z\right\rangle \left\langle z\right|_{a}.
\end{align}

The interaction Hamiltonian is cubic in the operators. To transform
this into differential form, it is convenient to transform to logarithmic
variables, and use basis operators $\hat{\Lambda}_{\eta}$ , $\hat{\Lambda}_{\phi}$
normalized in a logarithmic space defined so that:
\begin{align}
z & =e^{\eta}=e^{\eta'+i\eta"}\nonumber \\
\alpha & =e^{\phi}=e^{\phi'+i\phi"}.
\end{align}
 Using these variables, and introducing $m\left(\eta\right)=\left(3/2\right)\tanh\left(\eta'\right)$,
$n\left(\phi\right)=e^{2\phi'}-1$, one obtains simpler identities
such that:
\begin{align}
\hat{\sigma}_{z}\hat{\Lambda}_{\eta} & =2\left[\partial_{\eta}+m\left(\eta\right)\right]\hat{\Lambda}_{\eta}\nonumber \\
\hat{n}\hat{\Lambda}_{\phi} & =\left[\partial_{\phi}+n\left(\phi\right)\right]\hat{\Lambda}_{\eta}.
\end{align}
After making use of the differential identities, and transforming
to logarithmic variables, one obtains that $\partial_{\tau}\dot{Q}=iTr\left[\hat{\rho}\left[\hat{\sigma}_{z}\hat{n}\hat{\Lambda}-\hat{\Lambda}\hat{\sigma}_{z}\hat{n}\right]\right]/2$.
After substituting the identities, this gives a generalized Fokker-Planck
equation with constant diffusion, of the traceless form given in Eq
(\ref{eq:generalized FPE}): 
\begin{align}
\partial_{\tau}Q & =\left[\partial_{\eta"}n+\partial_{\phi"}m+i\left(\partial_{\phi}\partial_{\eta}-\partial_{\phi}^{*}\partial_{\eta}^{*}\right)\right]Q.
\end{align}
This has a similar behavior to the earlier gain equations, and leads
to a Q-function solution corresponding to Eq (\ref{eq:corr-cat-2}).
Therefore, the same general type of forward-backward stochastic equations
are obtained as previously. We note that there is a symmetric physical
explanation of each term. The drift term in the qubit phase, $\eta"$,
is driven by the meter occupation number $n$, while the drift term
in the meter phase, $\phi"$, is driven by the qubit occupation number
$m$. The remaining terms describe quantum noise, and are symmetric
in both the meter and qubit.

To analyze the observations or meter readings in detail, we define
a measured spin projection as the observed measurement result from
the meter, where the spin projection is inferred from the $x$ quadrature:
\begin{equation}
\sigma_{m}=\frac{q}{G}\,.
\end{equation}

\begin{figure}
\begin{centering}
\includegraphics[width=1\columnwidth]{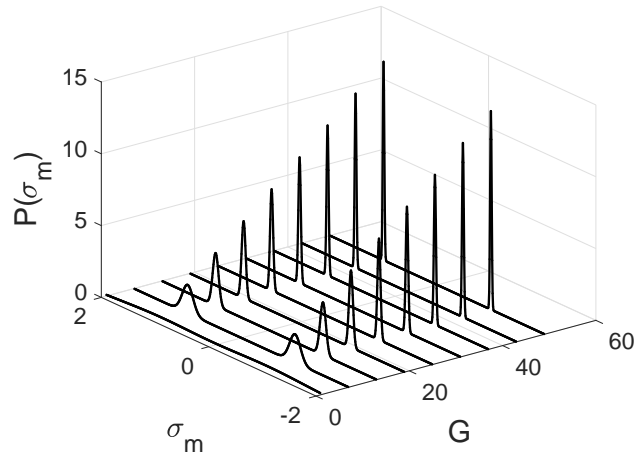} 
\par\end{centering}
\caption{Probability density $P\left(\sigma_{m}\right)$ for a spin measurement,
as a function of the measured spin projection $\sigma_{m}$ and measurement
gain $G$. The increasing sharpness with gain demonstrates the measurement
outcome for a discrete spin observable. The two peaks correspond to
the two possible spin projections. \label{fig:Inferred-spin-distribution}}
\end{figure}

Observations on the meter only require the output probability for
the stochastic field, which at time $\tau=\tau_{m}$ is given by:
\begin{align}
Q\left(\alpha,\tau_{m}\right) & =Tr\left[\hat{\rho}\left(\tau_{m}\right)\hat{\Lambda}_{b}\left(\alpha\right)\right]\nonumber \\
 & =\frac{1}{2\pi}\left[e^{-\left|G-\alpha\right|^{2}}+e^{-\left|G+\alpha\right|^{2}}\right].
\end{align}
The resulting probability distributions, as they develop in time,
are shown in Fig \ref{fig:Inferred-spin-distribution} for different
gains $G$. This plots $P(\sigma_{m})$, the probability of a measured
value $\sigma_{m}$ obtained from the Q-function for $x$, after integrating
over the transverse coordinate $p$, where: 
\begin{equation}
P\left(\sigma_{m}\right)=\frac{G}{2\sqrt{\pi}}\left[e^{-G^{2}\left|\sigma_{m}-1\right|^{2}}+e^{-G^{2}\left|\sigma_{m}+1\right|^{2}}\right]\,.
\end{equation}

We have shown that a Q-function phase-space coordinate has a distribution
that becomes relatively sharp after a high-gain measurement. This
allows an observer to determine which trajectory is the objectively
real one, and hence to calculate with a reduced phase-space ensemble.
This corresponds to an ``epistemic'', or information based, projection
of $\hat{\rho}$. For $G$ large, the two different values $\pm1$
for $\hat{\sigma}$ can also be measured by the different sign of
the outcomes for $\sigma_{m}$ or $q$, using a binning method to
reduce the effects of quantum noise.

\subsection{Bell inequality violations\label{subsec:Bell-inequality-violations}}

The strongest objections to realistic interpretations are through
correlated measurements that violate Bell inequalities, which have
been studied previously using both Q and P function methods \citep{rosales2014probabilistic,Drummond2014-bell-sim,ReidqubitPhysRevA.90.012111}.
We now consider correlated, spatially separated spins in the Bell
state \citep{Bell1964},
\begin{equation}
\left|\psi\right\rangle =\left(\left|\uparrow\right\rangle _{A}\left|\downarrow\right\rangle _{B}-\left|\downarrow\right\rangle _{A}\left|\uparrow\right\rangle _{B}\right)/\sqrt{2}.\label{eq:Bell state}
\end{equation}
 In the experiment, four different types of correlated measurements
are made for spins $\sigma_{\theta}^{A}$ and $\sigma_{\phi}^{B}$
in the $\theta$ and $\phi$ directions, and the results compared.
We assume that the eigenvalue relationships are such that:
\begin{align}
\hat{\sigma}_{\theta}^{A}\left|\uparrow\right\rangle _{\theta}^{A} & =\left|\uparrow\right\rangle _{\theta}^{A}\nonumber \\
\hat{\sigma}_{\theta}^{A}\left|\downarrow\right\rangle _{\theta}^{A} & =-\left|\downarrow\right\rangle _{\theta}^{A}.
\end{align}
The Clauser-Horne-Shimony-Holt Bell inequality is that, for a local
hidden variable model of quantum mechanics, given four different correlations
$E(\theta_{i},\phi_{j})=\left\langle \hat{\sigma}_{\theta_{i}}^{A}\hat{\sigma}_{\phi_{j}}^{B}\right\rangle $
of spin, one must have \citep{Clauser1969,clauser1978bell,brunner2014bell}:\textbf{
\begin{equation}
B=E(\theta_{1},\phi_{1})-E(\theta_{1},\phi_{2})+E(\theta_{2},\phi_{2})+E(\theta_{2},\phi_{1})\leq2\,.
\end{equation}
}

We can now use the stochastic Q-function for a calculation of these
four measured correlations, by extending the phase-space of the previous
subsection, now with four complex variables. Each different correlation
corresponds to a different measurement Hamiltonian, namely:
\begin{equation}
H\left(\theta,\phi\right)=\hbar g\left[\hat{\sigma}_{\theta}^{A}\hat{n}^{A}+\hat{\sigma}_{\phi}^{B}\hat{n}^{B}\right]\,.
\end{equation}

In this case, one has that the coupled initial state, including the
meter states at the two locations, is
\begin{equation}
\left|\psi\left(0\right)\right\rangle _{qm}=\frac{1}{\sqrt{2}}\left(\left|\uparrow\right\rangle _{A}\left|\downarrow\right\rangle _{B}-\left|\downarrow\right\rangle _{A}\left|\uparrow\right\rangle _{B}\right)\left|G/i\right\rangle _{A}\left|G/i\right\rangle _{B}.
\end{equation}
Here $G$ is the gain of each meter as previously. The final measured
state after evolution under a measurement time of $t_{m}=\pi/2g$
is:
\begin{eqnarray}
\left|\psi\left(t_{m}\right)\right\rangle _{qm} & = & \frac{1}{\sqrt{2}}\left|\uparrow\right\rangle _{A}\left|\downarrow\right\rangle _{B}\left|G\right\rangle _{A}\left|-G\right\rangle _{B}\nonumber \\
 &  & -\frac{1}{\sqrt{2}}\left|\downarrow\right\rangle _{A}\left|\uparrow\right\rangle _{B}\left|-G\right\rangle _{A}\left|G\right\rangle _{B}
\end{eqnarray}

\begin{figure}
\begin{centering}
\includegraphics[width=1\columnwidth]{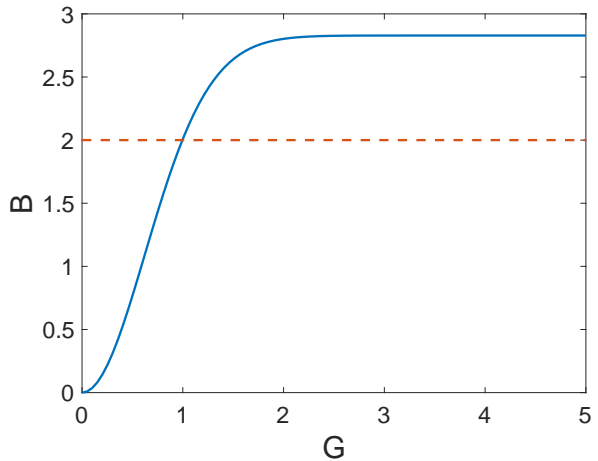}
\par\end{centering}
\caption{Measured Bell correlation $B\left(G\right)$ obtained from the Q function
for a spin-measurement outcome, using a binned inferred spin projection
$\sigma_{b}$ and measurement gain $G$. An inferred Bell violation
is obtained for gains $G\gtrsim1$. \label{fig:Inferred-Bell}}
\end{figure}

With the view that the phase space coordinates of the meter are elements
of reality, we can calculate the observed correlations using the procedure
outlined above for observing the spin orientation from the output
sign of the meter quadratures. The four correlations are obtained
from the inferred spins through a process of binning each phase space
coordinate to give a binary estimate of the measured spin, $\sigma_{b}=Sgn\left(\sigma_{m}\right)$.
We find that the Bell violation is a function of the gain, and is
given by $B=2\sqrt{2}\eta^{2}\left(G\right)$,\textcolor{red}{{} }where
$\eta\left(G\right)=\frac{1}{2}\left(erf(G)+1-erfc(G)\right)$ is
the binning efficiency, which is reduced at low gain. The resulting
Bell violation is plotted in Fig \ref{fig:Inferred-Bell}.

A Bell violation of $B>2$ is obtained for high enough efficiency,
and $G>1$ is already enough to observe this.

\subsection{Phase space and macroscopic realism}

Macroscopic interpretations can be gained from other quantum phase
space methods \citep{reid2017interpreting}. The Wigner function is
positive - in fact, it becomes the Q function - with the addition
of gaussian noise at the level of Planck's constant, which is macroscopically
negligible. This gives an interpretation for a macroscopic superposition
state, after interaction with an amplifying meter, in terms of a macroscopic
hidden variable, as explained in \citep{reid2016interpreting,reid2017interpreting}.
Consider a superposition of two position eigenstates, or a superposition
of two coherent states with real amplitudes $q_{1}$ and $q_{2}$,
so that a measurement of the position quadrature $\hat{q}$ gives
two macroscopically distinct outcomes, $q_{1}$ and $q_{2}$. The
existence of a positive phase space distribution means that a macroscopic
hidden variable exists, to determine the outcome of the measurement
of the quadrature $q$ to be either $q_{1}$ or $q_{2}$, to within
an uncertainty of order Planck's constant. The two outcomes are distinguishable
for $q_{1}$ and $q_{2}$ when macroscopically distinct. A similar
interpretation is given for the entangled cat-state 
\begin{eqnarray}
|\psi\rangle & = & \frac{1}{\sqrt{2}}\left[\left|\uparrow\right\rangle \left|\alpha\right\rangle +\left|\downarrow\right\rangle \left|-\alpha\right\rangle \right]\,.\label{eq:corr-cat-2-1}
\end{eqnarray}
where here $\alpha$ is large and real, and we take $q_{1}=\alpha$
and $q_{2}=-\alpha$. The Wigner or Q function for this state gives
a description that is consistent with the validity of a macroscopic
local hidden variable, that predetermines the outcome of the measurement
of quadrature $\hat{q}$ to be either $q_{1}$ or $q_{2}$, since
the non-positive nature of the Wigner distribution is small in the
macroscopic limit.

In this way, without decoherence, the system can be viewed as being
``in one state or the other'', where by ``states'' one simply
refers to a description predetermining the outcome to be either $q_{1}$
or $q_{2}$. Hence, one has an interpretation in which the system
is not paradoxically in ``both states at once'', which gives a partial
resolution of the measurement or Schr\"odinger cat paradox. A paradox
remains however, at the microscopic level (of order $\hbar$), in
that the system cannot be viewed as being in one or other of two \emph{quantum}
states, that determine the system to be $q_{1}$ or $q_{2}$ \citep{Reid2000PhysRevLett,reid2019criteria,Reid2018PhysRevA}. 

The reality model presented in this paper is consistent with the above
interpretation, but deepens and extends it, suggesting that the time-evolution
of the complex stochastic variable of the Q function corresponds to
a trajectory for an individual ``element of reality'' at all times.
Thus, for example, if the original quantum state in our measurement
example were a mixture or eigenstates of $q$ rather than a superposition,
one would see little difference in the $q$ trajectories under the
gain dynamics, but the $p$ trajectories would differ microscopically.
In this approach, the \emph{distribution} is related to knowledge
or epistemology, while the phase-space trajectory is ontological,
or \emph{real even at the microscopic level}. The epistemological
nature of the probability distribution means that an observer may
decide that a measurement gives them an improved knowledge of the
distribution. This Bayesian inference has no direct effect on the
objective trajectories. 

If there are reservoir couplings to the system, decoherence is introduced.
Quantum mechanics predicts that the reduced density matrix for a superposition
of eigenstates $|q_{1}\rangle$ and $|q_{2}\rangle$ evolves into
a mixture of $|q_{1}\rangle$ and $|q_{2}\rangle$. That is, the reduced
system-meter density matrix after decoherence corresponds to a probabilistic
mixture of the two eigenstates. The corresponding reduced $Q$ function
describes an ensemble of system-meter pairs. The trajectories of the
mixture differ from those of the superposition in that, while the
trajectories for $q$ are similar, the trajectories for the complementary
observable $p$ are different. However, in the standard quantum approach,
an additional non-unitary projection is needed to have just one outcome.
This is not necessary in our interpretation. A single macroscopic
outcome is a natural consequence of having one realistic trajectory
at all stages, both microscopic and macroscopic, and this requires
neither decoherence nor wave-function collapse.

\section{Quantum paradoxes\label{sec:Paradoxes}}

There are many apparent contradictions between what quantum mechanics
predicts and what one might expect in an ontological model. In this
section the most well-known paradoxes are discussed briefly. We explain
how these are resolved, at least in principle. Since the conventional
interpretation has a long history, we cannot treat all of the issues
here, although further detailed analysis will be carried out elsewhere.
Hence, the discussion given below is brief, with an emphasis on conceptual
points. We summarize possible objections and paradoxes in this interpretation
of quantum mechanics, and their resolution, as follows:

\subsection{Quantum fluctuations}

Since Q-functions are statistical, including vacuum fluctuations,
how can they represent eigenstates of measurements? In a full representation
of physical reality, the measuring device should be included. Examples
are treated above, and in each case the vacuum fluctuations are suppressed
relative to the measurement outcome, by a factor equal to the overall
gain. Thus, the final measurement outcome becomes sharp. It corresponds
precisely to the amplified value of the eigenvalue, after accounting
for the gain, for a system initially in an eigenstate of the observable
being measured.

\subsection{The Einstein-Podolsky-Rosen (EPR) paradox}

Given the EPR argument \citep{Einstein1935,Reid:2009_RMP81}, surely
this must be a local hidden variable theory? The ``elements of reality''
given in our model are not local hidden variables in the sense defined
by EPR. This is shown in Subsection \ref{subsec:Bell-inequality-violations}
by the violation of a Bell inequality. The dynamics of the Q-function
requires diffusive propagation in a \emph{backward} time direction.
Correlations permitted in retrocausal propagation are inconsistent
with EPR's local realism postulates, but they are allowed here. This
can be analyzed through the parametric interaction \citep{Reid:1989}
used to demonstrate the EPR argument, and will be treated elsewhere.

\subsection{Schr\"odinger's cat and entanglement}

How does this resolve Schr\"odinger's question \citep{schrodinger1935gegenwartige},
that a state such as (\ref{eq:corr-cat-2}) after the measurement
suggests a macroscopic object in two places $q_{1}$ and $q_{2}$
at once, like a cat simultaneously ``living and dead''? This issue
arises in the measurement problem, where the measured system becomes
entangled with the meter, as we describe in Section\textcolor{red}{{}
}\ref{sec:Idealized-models-of}. The explanation is that the Q function
evolves as the system interacts with the measurement device. The scaled
projection along $p$ at any given time is plotted in Fig (\ref{fig:Inferred-spin-distribution}).
At high enough gain, so that $q_{1}$ and $q_{2}$ are macroscopically
separated, this gives a definite result compatible with quantum predictions,
and consistent with the pointer being \emph{either} at $q_{1}$ or
$q_{2}$, as explained in \citep{reid2016interpreting,reid2017interpreting}.
The cat is either living or dead, but not both.

\subsection{Quantum to classical transition}

As described in Section \ref{sec:Idealized-models-of} and above,
our model for reality gives an explanation of the Schr\"odinger cat
paradox. The superposition state of Eq \ref{eq:corr-cat-2-1} is consistent
with macroscopic reality, since for large $\alpha$, the result of
the measurement of $\hat{x}$ is for any individual system given by
a well-defined stochastic field quadrature value. Cat-states of this
type arise as a result of the measurement process, which amplifies
microscopic superposition states into macroscopic (entangled) ones.
This occurs when a macroscopic device is coupled to a smaller system.
Our model provides an explanation of the measurement problem in which
the quantum-to-classical transition takes place via amplification,
rather than decoherence.

Decoherence is significant in the longer term, because any coupling
to a reservoir introduces loss of information about the entangled
system. This brings about an irreversible dynamics. In our model,
however, macroscopic realism holds prior to decoherence.

This picture also explains why there is such sensitivity of a macroscopic
superposition state to a very small amount of decoherence. It is known
that the macroscopic superposition state collapses more quickly, or
is more sensitive to decoherence, as the size of the separation $\left|q_{1}-q_{2}\right|$
increases \citep{brune1996observing}. The explanation of this in
terms of the macroscopic reality interpretation given by phase-space
methods, and also in terms of the ``elements of reality'' of the
cat-states, was suggested in \citep{reid2016interpreting,reid2017interpreting}.
However, our detailed microscopic model gives a more precise explanation
in terms of trajectories. As $\alpha$ increases, the phase space
trajectories for $q$ become indistinguishable from those of a classical
mixture of the states $\left|q_{1}\right\rangle $ and $\left|q_{2}\right\rangle $.
Differences occur for the trajectories in the complementary observable
$p$, but these become close to that of the mixture, as $\alpha$
increases. Thus the sensitivity to decoherence can be viewed as the
underlying dynamics for the quantum state becoming close to that of
the mixture, as $\alpha$ increases, and hence more readily disrupted
by the reservoir couplings.

\subsection{Bell's theorem and causality}

Bell's theorem \citep{Bell1964} proves that if one attempts to complete
quantum mechanics with local hidden variables, there will be a contradiction
with quantum experiments \citep{Greenberger1989,Mermin1990-entanglement,Ardehali1992,Belinsky1993-N-particle,Collins2002,Svetlichny1987}.
We have shown that Bell violations are obtained using phase-space
variables as elements of reality, if one includes the measurement.
It is already known from absorber theory \citep{tetrode1922causal,wheeler1945interaction,pegg1982time},
that future time boundary conditions can cause violation of Bell inequalities.
A similar backward time propagation occurs in Q-function dynamics
\citep{altland2012quantum}. One can have locality, with no information
transfer \citep{hillery1995bell}, while violating the Bell inequality.
In the multipartite case, up to 60 simultaneous spatially separated
measurements with genuine multi-partite entanglement and non-locality
have been simulated using these methods \citep{ReidqubitPhysRevA.90.012111}.
Due to vacuum fluctuations, a model of the measurement is needed for
Bell violations and ``all-or-nothing'' effects \citep{greenberger1989going}.

\subsection{Particle statistics}

Husimi Q-functions are defined for bosons, yet the physical world
includes fermions. Research on fermionic Gaussian operators \citep{Corney_PD_JPA_2006_GR_fermions}
shows that fermionic Q-functions exist \citep{FermiQ,joseph2019entropy}
with a real and positive distribution. This complete phase space is
based on the bounded homogeneous spaces of group theory \citep{cartan1935domaines,Hua_Book_harmonic_analysis}.
Any fermionic operator expectation value and its dynamics \citep{Riashock2018}
can be computed, in a similar way to the Husimi Q-function. While
we have used simpler SU(2) coherent state projectors in this paper,
these can be generalized to Fermi fields.

\subsection{Uniqueness}

Why should coherent state projectors represent macroscopic realism,
as opposed to other measurement operators? Ontological models should
not depend on how measurements are implemented. Coherent state projectors
provide a minimal, unbiased implementation of the group symmetry of
the field commutators in the standard model. Their role was recognized
by Schr\"odinger \citep{Schrodinger_CS} and others \citep{Glauber1963-states}
who proved that they have classical behavior at the macroscopic level.

\subsection{Delayed choice}

The measurement problem is also related to delayed choice experiments
\citep{wheeler1978past}. This idea was proposed by Wheeler, who was
also responsible (with Feynman) for absorber theory, which was an
early attempt at understanding future boundary conditions in electrodynamics.
The fact that boundary conditions in the future propagate retro-causally
in amplifiers provides an elegant explanation of why choice of measurement
can be made at a later time. This will be treated in detail elsewhere.

\subsection{Contextuality}

Because the Q-function model gives the same predictions as quantum
mechanics, it has the same contextuality properties \citep{kochen1967problem}.
It achieves this because the model includes interactions with measuring
devices. This was also the nature of Bohr's response to the EPR paradox
\citep{Bohr1935CanQuant}. When there is a new context, that is, a
different measuring device, the measured result is changed through
retrocausal interactions. This is completely different to that occurring
in a classical, hidden variable model, of the type that is used to
analyze contextuality in the Kochen-Specker theorem. Contextuality
is not really unexpected in a model of reality, since the context
is part of the physical world. 

\section{Summary}

We have shown that a realistic ontological model for quantum mechanics
is obtainable by utilizing a phase-space of stochastic fields. The
Q-function gives the probability of a given field configuration as
a generalized phase-space coordinate. Vacuum fluctuations are suppressed
in measurements with gain, for both continuous and discrete eigenvalue
examples. After the gain is included, sharp eigenvalues are obtained.
Thus, the fields can represent an objective physical reality, without
an explicit collapse on measurement. 

The coherent states are complete and give a unique set of ``addresses''
in the quantum world, with positive probability. There is also an
exact mapping from unitary quantum field dynamics to Q-function dynamics,
which is obtained from the operator identities. This uses the fact
that quantum field Hamiltonians are at most quartic in the quantum
fields. Quantum dynamics can therefore be re-expressed as differential
equations that have a time-symmetric action principle and path integral.

Our discussion of measurement focuses on idealized cases, just as
in most discussions of measurement, because the goal is to understand
how the intrinsically noisy Q-function distribution can describe an
idealized measurement, leading to a well-defined eigenvalue. One can
certainly include noise and decoherence to make even more realistic
models. We do not do this here, as it tends to obscure the fundamental
issue, which is that measurement involves gain or amplification. Gain
provides a mechanism by which intrinsic microscopic vacuum noise is
relatively suppressed.

The main thrust of this paper was to show how the inclusion of a meter
allows one to obtain well-defined eigenvalues in amplified measurements.
This is an important fundamental issue, since the stochastic fields
themselves have quantum fluctuations, and a well-defined eigenvalue
is not compatible with large fluctuations in the measured result.
Q-function dynamical equations, as we have shown, give exactly the
required properties. Quantum noise is reduced through retro-causal
propagation in the amplified quadrature of a low-noise measuring device.
Since only one of the two complementary quadratures of a field can
be amplified in this way, the physics of measurement enforces the
uncertainty principle.

The Q-function dynamical equations for unitary evolution have a traceless
diffusion, with equal positive and negative diffusion terms, leading
to retrocausal effects from boundary conditions in the future. Related
phenomena in electrodynamics were studied by Tetrode, Dirac, Wheeler
and Feynman \citep{tetrode1922causal,wheeler1945interaction}, which
Dirac called a ``beautiful'' aspect of his theory. Here, negative
diffusion terms occur in parametric amplifier equations, and are directly
responsible for the sharp measurement results given above. As a result,
Bell's arguments about local hidden variable theories do not apply
to Q-functions. The model has a different type of time-evolution to
a hidden variable theory. One can study more complicated effects,
and include decoherence if necessary, but this is not essential to
the development of measurement results: the only required ingredient
is the meter itself, as indeed one might expect physically.

In summary, we demonstrate the existence of an objective, relativistically
invariant ontology for quantum fields.
\begin{acknowledgments}
PDD and MDR thank the generous hospitality of ITAMP at Harvard University,
the Weizmann Institute of Science, and the Joint Institute of Laboratory
Astrophysics of The University of Colorado. This work was funded through
Australian Research Council Discovery Project Grants DP180102470 and
DP190101480, and performed in part at Aspen Center for Physics, which
is supported by National Science Foundation grant PHY-1607611. 
\end{acknowledgments}

\section*{Appendix: interpretations of quantum measurement }

There are a variety of previous interpretations of quantum mechanics
which are reviewed in several places \citep{bell2004speakable,elitzur2006quo,schlosshauer2005decoherence}.
Without giving a complete survey, we will provide a compressed summary
of most earlier approaches, and explain how they differ from our proposal. 

We will start with Leggett's classification of three common strands
in quantum measurement interpretations \citep{leggett2005quantum},
noting that some interpretations combine more than one of these ideas.
These three common ideas are the statistical interpretation, many
worlds, and the decoherence approach. We then consider three approaches
which add additional structure to the usual Hilbert space, namely
the Bohm theory of pilot-waves, the spontaneous collapse idea, and
an alternative phase-space interpretation.

\subsection{The~statistical~interpretation}

The original Copenhagen theory of measurement proposed that quantum
mechanics only gives information about the statistics of measurements
on an ensemble of identically prepared experiments \citep{ballentine1970statistical,born1955statistical}.
Thus, it provides no ontological interpretation of the microscopic
events that cause the statistics. The founder of this approach was
Bohr, whose measurement interpretation is in almost all standard textbooks
\citep{dirac1981principles}. Several later viewpoints share this
approach of a statistical interpretation. The most well-known are
QBism \citep{fuchs2014introduction} and consistent histories \citep{griffiths1984consistent},
although they have other features as well. In its original form, the
statistical interpretation is unable to explain how a measurement
differs from ordinary unitary evolution, and what is the underlying
reality that the statistical theory describes. It is particularly
silent on what happens between measurements, or in other words, ``is
the moon there when nobody looks?'' \citep{mermin1985moon}. 

\subsection{The many-universe interpretation}

This interpretation \citep{Everett:1957} regards the Hilbert space
state-vector as objectively real. It sees no contradiction in the
proposal that the universe `really' is a macroscopic superposition
of many possibilities. In this approach observers are not aware of
other universes. The explanation of how and why observers become unaware
of all the other universes, and which of these are to be regarded
as 'real', is a difficulty with many universe theory. This approach
has been criticized as not properly accounting for space-time properties
\citep{wallace2010quantum}. A variant of this uses multiple interacting
classical universes \citep{HallPhysRevX}, although in a restricted,
non-relativistic context. 

What we propose is a model of reality as a \emph{single} universe.
The fact that many realizations are possible is a different thing
to claiming that they all exist simultaneously. In this sense our
proposal is akin to statistical mechanics, which also allows for many
possible realizations, while only requiring one to exist. A theory
of one universe is more parsimonious than a many-universe theory.
A picture of just one universe seems preferable according to Occam's
razor \citep{schaffer2015not}, as requiring fewer assumptions. It
leads to a picture of reality as objective fields that exist in space-time,
which is a simpler physical understanding of ``what is''.

\subsection{The decoherence interpretation}

This approach focuses on the fact that macroscopic states decohere,
or become highly entangled with their environment. As Bell \citep{bell2004speakable}
and Leggett \citep{leggett2005quantum} have emphasized, this still
leaves open the question of how just one of the resulting macroscopic
states emerges as the preferred alternative during measurement. Decoherence
interpretations do not provide a mechanism to project out an individual
eigenstate. This requires a non-unitary step in conventional quantum
mechanics. As a result, Leggett, Bell and others have questioned whether
decoherence actually solves the measurement problem. 

The projection issue is apparently avoided if decoherence is combined
with a many-universe interpretation, as in Quantum Darwinism \citep{Zurek1981,Zurek_PhysRevA.73.062310},
and other related approaches. It might be argued that this approach
leads to a lack of parsimony, pointed out above, and there are known
technical objections \citep{kastner2014einselection} as well.

In our stochastic field approach, decoherence also occurs whenever
a system is in contact with a reservoir. This is a normal part of
quantum evolution. However, the question of whether an observable
is macroscopic, that is, large enough for a macroscopic observer to
register it, requires amplification. \emph{Therefore, we regard the
amplification step as fundamental to measurement. }Decoherence occurs
in many situations, but gain is essential to having macroscopic or
'classical' measurement results, as pointed out by Bohr in his discussions
with Einstein at Solvay \citep{bohr1996discussion}. In the Q-function
model, decoherence will create irreversibility in a measurement, if
it is present, but it is not required in order to explain having a
definite outcome.

\subsection{Pilot-wave theory}

In the Bohm theory \citep{BohmPhysRev.85.166}, particles all move
in a potential that depends on the wave-function. This formulation,
in its original version, has no quantum fields. While there are extensions
of this approach to treat quantum field theories \citep{struyve2010pilot},
not all aspects of quantum fields can be treated in a pilot-wave approach
as yet. The stochastic field approach we propose is directly based
on quantum field theory, and uses fields as its ontological elements.
It has no wave-function requirement, since the Q-function dynamics
can be worked out without having a wave-function. A Hilbert space
or quantum state can be used if convenient, but is not fundamental. 

\subsection{Spontaneous collapse proposals}

Another proposed route to solve measurement problems is the use of
additional mechanisms outside of quantum mechanics. These may involve
either a random collapse with no currently known physical origin \citep{ghirardi1986unified},
or a mechanism involving gravity \citep{diosi1987universal,penrose1996gravity}.
In general, the hallmark of these approaches is that there is an extra
parameter with an unknown magnitude. While such processes may exist,
they are not necessary in the present interpretation of reality. Extensions
to treat additional decoherence including gravity in a Q-function
realism model are certainly possible. However, they do not appear
necessary.

\subsection{Other phase-space ontologies}

There is another phase-space interpretation of quantum measurement
\citep{budiyono2017quantum}, using a different approach. This uses
non-relativistic quantum mechanics, and considers ontic variables
with an epistemological restriction of the order of Planck's constant.
It has distributions rather than actual trajectories as elements of
realism. Since it is not based on fields, it is unclear whether it
can reproduce the levels of current precision of quantum field theory.
Hydrogen-like spectra are calculated and measured to twelve decimals
\citep{beyer2013precision}, and helium-like spectra to six decimals
\citep{Chantler2012PhysRevLett}, all of which require a full theory
of QED for their accurate prediction. 

\bibliographystyle{apsrev4-1}
\bibliography{Reality,RealitySup}

\end{document}